\documentclass[twocolumn]{aastex631}

\usepackage{slashbox,pict2e}
\usepackage{diagbox}

\newcommand{\de}{\mathrm{d}}
\newcommand{\jet}{\mathrm{j}}

\def\s{\mathrm{s}} %.......seconds
\def\m{\mathrm{m}} %...........meters
\def\cm{\mathrm{cm}} %........centimeters
\def\km{\mathrm{km}} %........kilometers
\def\g{\mathrm{g}} %...........grams
\def\erg{\mathrm{erg}} %.......ergs
\def\J{\mathrm{J}} %...........joules
\def\K{\mathrm{K}} %...........Kelvin

\def\G{{\rm G}} %...........gauss
\def\T{{\rm T}} %...........tesla
\def\ej{\mathrm{ej}}

\def\tot{\mathrm{tot}}     %...total
\def\iso{\mathrm{iso}} % isotropic
\def\stop{\mathrm{stop}} % stop
\renewcommand{\epsilon}{\varepsilon}

\newcommand{\engine}{\mathrm{e}}
\newcommand{\te}{t_\mathrm{e}}
\newcommand{\td}{t_\mathrm{d}}
\newcommand{\esc}{\mathrm{esc}}
\newcommand{\Light}{\mathrm{L}}

\newcommand{\crit}{\mathrm{crit}}

\begin{document}

\title{The collimation of relativistic jets in post-neutron star binary merger simulations}

\author[0000-0002-4243-3889]{Matteo Pais}
\affiliation{Racah Institute for Physics, The Hebrew University, Jerusalem 91904, Israel}

\author[0000-0002-7964-5420]{Tsvi Piran}
\affiliation{Racah Institute for Physics, 
The Hebrew University, 
Jerusalem 91904, Israel}

\author[0000-0002-2316-8899]{Yuri Lyubarsky}
\affiliation{Physics Department, 
Ben-Gurion University of the Negev,
POB 653, Be'er-Sheva 84105, Israel}

\author[0000-0003-4988-1438]{Kenta Kiuchi}
\affiliation{Max Planck Institute for Gravitational Physics (Albert Einstein Institute) \\ 
Am Mühlenberg 1, 14476, Potsdam-Golm, Germany}

\affiliation{Center for Gravitational Physics and Quantum Information \\
Yukawa Institute for Theoretical Physics, Kyoto University,
606-8502, Kyoto, Japan}

\author[0000-0002-4979-5671]{Masaru Shibata}
\affiliation{Max Planck Institute for Gravitational Physics (Albert Einstein Institute) \\ 
Am Mühlenberg 1, 14476, Potsdam-Golm, Germany}

\affiliation{Center for Gravitational Physics and Quantum Information \\
Yukawa Institute for Theoretical Physics, Kyoto University,
606-8502, Kyoto, Japan}

%%%%%%%%%%%%%%%%%%%%%%%%%%%%%%%%%%%%%%%%%%%%%%%%%%

\begin{abstract}
The gravitational waves from the binary neutron star merger GW170817 were accompanied by a multi-wavelength electromagnetic counterpart, which confirms the association of the merger with a short gamma-ray burst (sGRB). 
The afterglow observations implied that the event was accompanied by a narrow, $\sim 5^\circ$, and powerful, $\sim 10^{50}$ erg, jet.
We study the propagation of a Poynting flux-dominated jet within the merger ejecta (kinematic, neutrino-driven and MRI turbulence-driven) of a neutrino-radiation-GR-MHD simulation of two coalescing neutron stars.
We find that the presence of a post-merger low-density/low-pressure polar cavity, that arose due to angular momentum conservation,  is crucial to let the jet break out. 
At the same time the ejecta collimates the jet to a narrow opening angle. 
The collimated jet has a narrow opening angle of $\sim 4$--$7^\circ$ and an energy of $10^{49}$--$10^{50}~\erg$, in line with the observations of GW170817 and other sGRBs. 
\end{abstract}


%%%%%%%%%%%%%%%%%%%%%%%%%%%%%%%%%%%%%%%%%%%%%%%%%%

\section{Introduction} \label{sec: intro}

The predicted association of short gamma-ray bursts (sGRBs) with binary neutron star (BNS) mergers \citep{eichler_nucleosynthesis_1989} initially had just indirect evidence. 
The most remarkable of those was the tentative observations of a kilonova following the sGRB 130603B \citep{tanvir_kilonova_2013,berger_r-process_2013}. Later on  a few other similar marginal events were detected  in other sGRBs \citep[e.g.,][]{  yang_possible_2015,jin_macronova_2016,jin_light_2015, gompertz_diversity_2018, ascenzi_luminosity_2019,  jin_kilonova_2020, lamb_short_2019, rossi_comparison_2020, fong_broadband_2021}. In 2017, 
GW170817, the first detection of gravitational waves from a BNS merger \citep{abbott_gw170817_2017} was followed by a sGRB \citep[e.g.,][]{abbott_multi-messenger_2017,goldstein_ordinary_2017, savchenko_integral_2017}. 
At first sight, it seemed that the long standing prediction has been confirmed. 

However, a closer examination of GRB 170817A revealed that it was not a regular sGRB. 
Its total isotropic equivalent energy ($\sim 10^{46}$) is smaller by three orders of magnitude than the weakest sGRB measured so far \citep{gottlieb_cocoon_2018} and by four orders of magnitude than typical sGRBs \citep{nakar_short-hard_2007}. 
Compactness arguments \citep{kasliwal_illuminating_2017, matsumoto_constraints_2019} revealed that the observed $\gamma$-rays could not have emerged from a regular GRB viewed off axis (with the difference in luminosity and hardness arising from a different Lorentz boost). 
It also showed that the $\gamma$-rays must have been produced in a mildly relativistic outflow with a Lorenz factor of $\Gamma\sim 2$--$3$ and conditions drastically different from those characteristic for regular GRBs.

Furthermore, unlike other sGRBs, both the radio \citep{hallinan_radio_2017} and X-ray \citep{troja_x-ray_2017} afterglows were detected only several days after the prompt $\gamma$-rays and they were initially much weaker than typical sGRB afterglow. 
Thus, based on the prompt $\gamma$-rays and the early afterglow one could question whether GW170817 was accompanied by a regular sGRB.   

Late observations of the afterglow of GW170817 enabled us to answer this question. 
The multi-wavelength afterglow peaked around 175~days after the burst. 
The shape of the light curve around the peak flux, as well as the apparent superluminal motion detected in VLBI observations \citep{mooley_superluminal_2018, mooley_optical_2022, ghirlanda_compact_2019}, revealed that this late-time emission was dominated by a very energetic ($ \sim 10^{49-50}~\erg$),  narrowly collimated ($< 5^\circ$) jet  that was observed from a viewing angle of about $20^\circ$ \citep{nakar_electromagnetic_2019}. 
The jet  emerged successfully from the ejecta surrounding the merger and most likely it produced a regular sGRB pointing along its axis.
This last point is not trivial as, even though the merger ejecta contains only a few percent of a solar mass, this may be sufficient to choke a powerful jet (see e.g. \cite{bromberg_are_2011} for jet propagation in a static matter and \cite{hamidani_jet_2019} and \cite{gottlieb_propagation_2022}  for jet propagation in an expanding  surrounding matter, as is the case in a merger). 
A related question is, for a powerful jet that manages to escape, how is it collimated  to such a narrow angle?  Our goal here is to address these two questions.

In this work we aim to determine the conditions that allow a jet to propagate in the post-BNS merger ejecta and successfully break out from it and, at the same time, how the ejecta affects the jet opening angle. 
To do so we analyze the results of a realistic general relativistic neutrino-radiation magneto-hydrodynamical (MHD) simulation of a one-second-long BNS merger presented in \cite{kiuchi_self-consistent_2022}.
The paper is structured as follows. 
We begin, in \S~\ref{sec: simulation}, with a brief description the setup of the  neutrino-radiation-GRMHD simulation of a BNS merger referring the reader to \cite{kiuchi_self-consistent_2022} for details. 
We outline, in \S~\ref{sec: theory},  the main theoretical estimates concerning the conditions required for a jet to escape from the ejecta and to be collimated by the external pressure. 
In \S~\ref{sec: results} we scrutinize the results of the post-merger phase of the BNS simulation.  Using 2D slices of the simulation box, we calculate the conditions for both the collimation and the escape of the jet over time. We  discuss the implications of our findings for GW170817 and other sGRBs in \S~\ref{sec:implications} and summarize our results in \S~\ref{sec:conclusions}.

%%%%%%%%%%%%%%%%%%%%%%%%%%%%%%%%%%%%%%%%%%%%%%%%%%

\section{Simulation} \label{sec: simulation}

The simulation for BNS mergers is performed using a code described in \citet{kiuchi_implementation_2022}, and
the simulation result of a BNS merger here used is reported in \citet{kiuchi_self-consistent_2022} in detail. 
The neutron star is modeled with the SFHo equation of state \citep{steiner_core-collapse_2013}. 
The binary is composed of $1.2~\mathrm{M}_\odot$ and $1.5~\mathrm{M}_\odot$ neutron stars. 
The corresponding chirp mass is consistent with the one observed in GW 170817.  
The simulation domain is composed of 13 levels of the fixed mesh refinement (FMR). 
The finest FMR domain has the size of $L\in[-37.875,37.875]~\km$ with the grid spacing of $\Delta x=150~{\mathrm m}$, while the coarsest has a grid size of $L\in[-1.55136,1.55136] \times 10^{5}~\km$ with a grid spacing of $\Delta x=614.4~\km$. 

The simulations that are general relativistic neutrino-radiation MHD last for one second following a BNS merger \citep{kiuchi_self-consistent_2022}. 
These are the longest simulations for BNS mergers carried out so far. 
The long time scale and the inclusion of neutrino and MHD physics make it possible to explore both the dynamical ejecta and the post-merger ejecta driven by the MHD power from the remnant disk as well as the density profile of the merger remnant around the black-hole spin axis. 
All these ingredients are critical for this work.  
The merger and the subsequent black hole formation take place at $\approx 0.015~\s$ and $\approx 0.032~\s$, respectively.

In \cite{kiuchi_self-consistent_2022}, the strong Poynting flux which could induce a relativistic jet is not found until the termination of the simulation because of the limited grid resolution (which results in the spurious underestimation of the black hole spin and the Poynting flux generated by the Blandford-Znajek mechanism) and/or a short computational time following the merger remnant. 
However, along the spin axis of the black hole, a funnel region with a high magnetization, which could eventually generate a relativistic outflow in the presence of an intense Poynting flux, is formed in the late stage of the remnant with $t \agt 1~\s$. 
In the following, we analyze the funnel region and explore the possibility of a jet formation in a hypothetical but plausible value of the Poynting flux. 

%%%%%%%%%%%%%%%%%%%%%%%%%%%%%%%%%%%%%%%%%%%%%%%%%%

\section{Theory} \label{sec: theory}

As the jet propagates through the expanding ejecta it
dissipates its energy in shocks that form at the jet head \citep[e.g.,][]{matzner_supernova_2003,Lazzati_universal_2005,bromberg_are_2011}. 
If the jet is not powerful enough it may not break out from the ejecta \citep[e.g.,][]{bromberg_are_2011,gottlieb_propagation_2022}. The shocked material 
forms a high-pressure bubble (the cocoon) that engulfs the jet and collimates it \citep[e.g.,][]{komissarov_magnetic_2009, lyubarsky_asymptotic_2009, bromberg_are_2011}. In the following paragraphs we explore the conditions within the merger ejecta and compare with those needed for break-out and for collimation.

\begin{figure*}
    \centering
    \includegraphics[scale=0.35]{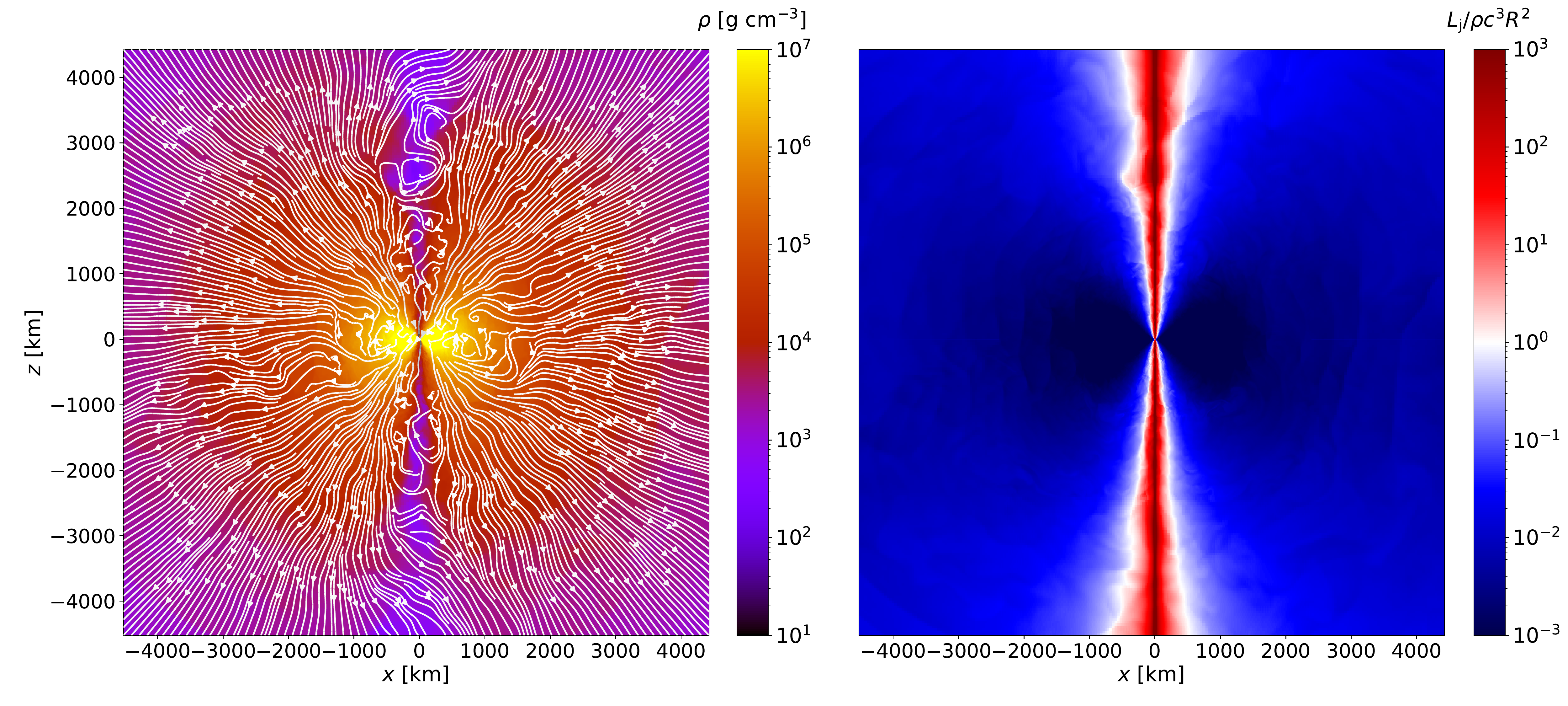}
    \caption{\emph{Left}: Density map in a logarithmic scale of a 2D slice ($x$-$z$ plane) through the center of the simulation box at $t=1~\s$. We superposed the velocity streamlines (white colour). 
    \emph{Right}: The jet parameter $\tilde{L}$  calculated for the 2D slice of the ambient density for  a jet luminosity of $10^{50}~\erg/\s$. Note the almost empty cone along the axis.}
    \label{fig: density_and_Ltilde}
\end{figure*}

\subsection{Escape} \label{subsec: escape}
To produce a GRB the jet must  break out from the moving ejecta. Recently \citet{gottlieb_propagation_2022} worked out the criteria for a jet break-out in such a case (see their Table~1). This criteria that was obtained for a uniform outflow and a top hat jet is based on a comparison of the kinetic energy of the jet with the kinetic energy of the surrounding outflow:
\begin{equation}
\label{eq: escape}
    E_{\jet,\rm{iso,tot}} > 150 [20]\left[\left( \frac{\td}{\te} \right)^2 + 2 \right] E_{\ej,\rm{iso,tot}}\theta_\jet^2 \ ,
\end{equation}
where $E_{\ej,\rm{iso,tot}}$  is the ejecta isotropic equivalent total energy and $E_{\jet,\rm{iso,tot}}$ and $\theta_\jet$ are the jet isotropic equivalent energy and opening angle. $\te$ and  $\td$ are the jet engine working time and  the delay time between the BNS merger (the onset of the mass ejection) and the jet launch, respectively.  The pre-factor 150 corresponds to baryonic jets while the factor 20 corresponds to magnetic-dominated jets. 

As the ejecta resulting from the merger is highly anisotropic we translate this condition  to our configuration assuming that the strong density gradients (see Fig.~\ref{fig: density_and_Ltilde}) do not influence significantly the jet propagation.  We do so by identifying the isotropic equivalent energy, $E_{\rm ej,tot} $, at $\theta_\jet$ as 
$E_{\ej,\iso,\tot}(\theta_\jet) = 4 E_{\ej}(< \theta_\jet)/\theta_\jet^2$, where $E_{\ej}(< \theta_\jet)$ is the integrated kinetic energy of the ejecta from the axis up to $\theta_\jet$. For a double sided jet  and assuming    $\te > \td$ from Eq.~(\ref{eq: escape}) we have:
\begin{equation}
\label{eq: escape2}
    E_{\jet } > E_{\jet, \esc} \equiv 300 [40] E_{\ej}( < \theta_\jet)\times \theta_\jet^2 \ ,
\end{equation}
where again the factor 300 corresponds to a hydrodynamic jet while the factor 40 to a magnetic one. 

\subsection{Collimation}

\begin{figure*}
    \centering
    \includegraphics[scale=0.36]{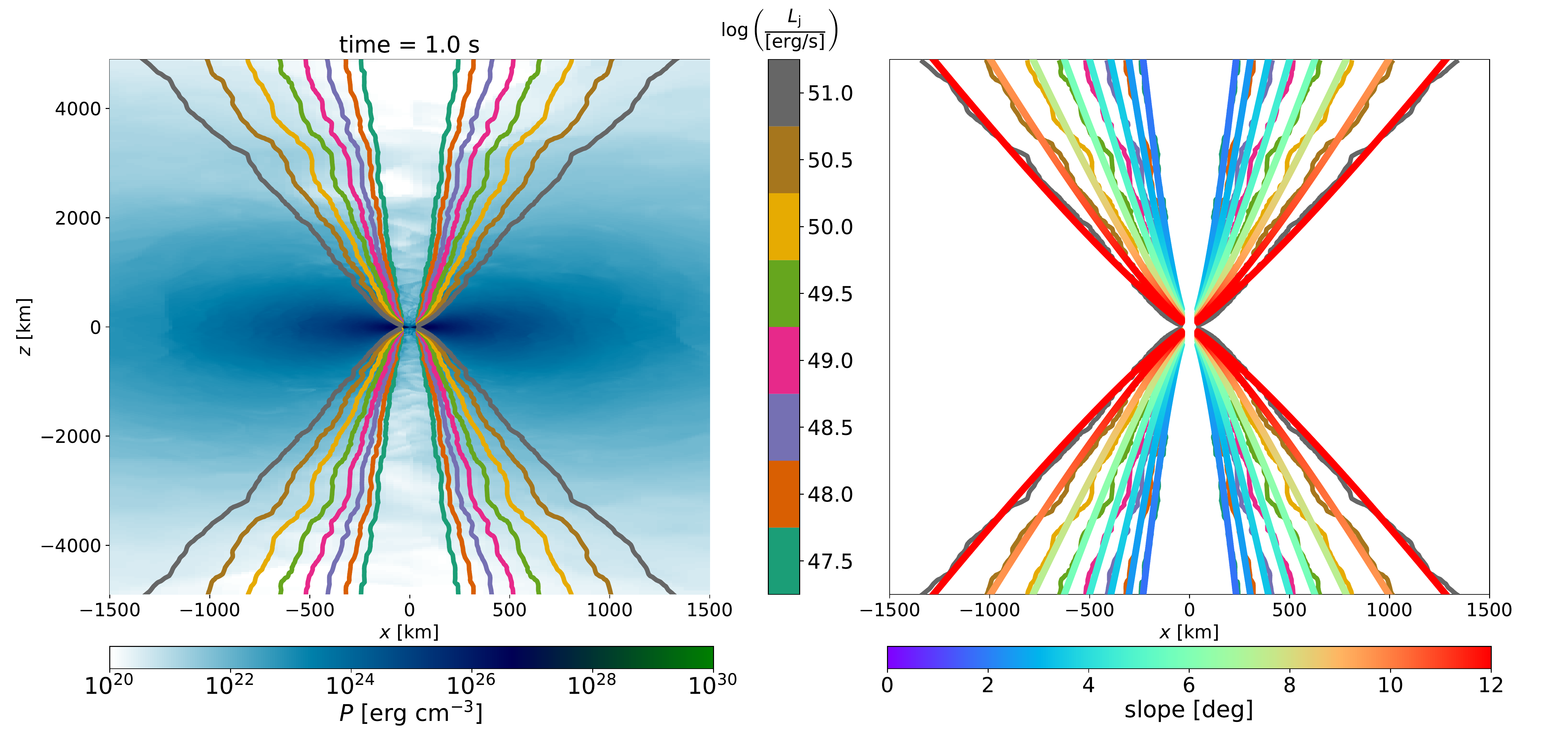}
    \caption{\emph{Left}: Map of the thermal pressure superposed onto the collimation contours that satisfy Eq.~(\ref{eq: pressure}). The bottom color bar shows the values of the ambient pressure while the different colors of the contour lines represent  different jet luminosities (vertical color bar in a log scale). We fixed the light cylinder radius at $R_\Light = 40~\km$. We chose a different scale ratio for the $x$ and $z$ axes to stretch the central under-dense polar region of the jet where the contours crowd. \emph{Right}: Same as the left panel but superposed onto a power law fit $z = a x^b$. The horizontal color bar at the bottom of the figure represents the local slope of the curve fit (in degrees). The curves with $L_\jet=10^{49}\erg/\s$ and $10^{50}\erg/\s$ show an outer slope between $4^\circ$ and $7^\circ$. 
    Note that the horizontal axis covers $x \in [-1500 , 1500]\,\km$ while the vertical axis covers $ z \in [-4500 , 4500]\,\km$ to better distinguish the curves closer to the central axis.
    }
    \label{fig: pressure_and_slope}
\end{figure*}

It is now generally accepted that highly relativistic outflows are launched hydromagnetically. 
It is assumed that the magnetic field is amplified in the accretion flow and the magnetic field lines tread the horizon of the rapidly rotating black hole. 
The rotation twists the magnetic field lines producing a Poynting dominated outflow.
The magnetic field in Poynting dominated outflows is spiral: each turn of the central body adds to the flow one more magnetic loop. 
In an expanded flow, the conservation of the magnetic flux implies that the poloidal magnetic field decreases as 
\begin{equation}
\label{eq: Bp}
B_\varrho= B_0\left(\frac{R_\Light}{R}\right)^{2},
\end{equation}
where $R$ is the cylindrical radius of the jet, $R_\Light$ the light cylinder radius, $B_0$ the magnetic field at the light cylinder. Then the azimuthal component of the magnetic field varies as 
\begin{equation}
\label{eq: Bphi}
B_{\phi}= B_0\frac{R_\Light}{R} \ .
\end{equation}
The jet luminosity is estimated as
\begin{equation}
\label{eq: luminosity}
   L_\jet=\frac 14 R_\Light^2B_0^2c \ .
\end{equation}

The outflow is collimated by the pressure of the confining medium. An important point is that when the jet escapes from the confining medium, it expands freely so that the collimation angle increases by $\sim\gamma^{-1}$, where $\gamma$ is the Lorentz factor of the flow at the escape point. 
This implies that the flow remains highly collimated only if it acquires a large enough Lorentz factor before it escapes. 
However, the flow is accelerated only when it laterally expands.
Therefore the confining medium should be extended enough; then the jet could expand so that the opening angle remains small.

 Because there is a nearly empty funnel along the axis of the ejecta, the collimation mechanism  of the jet differs from the sometimes considered collimation by the cocoon  \citep{bromberg_are_2011}. The cocoon is formed when the jet pushes its way through the external medium so that the matter ahead of the jet passes the bow shock, and then flows aside forming a high-pressure cocoon. The efficiency of the process is described by the parameter
\begin{equation}
\label{eq: Ltilde} 
    \tilde{L} = \frac{L_\jet}{\Sigma_\jet \rho_\mathrm{a} c^3 } \ , 
\end{equation}
where $\Sigma_\jet = \pi R^2$ is the cross section of the jet at a given radius, $\rho_\mathrm{a} (R,z)$  the local ambient density, and $L_\jet$ is the hypothetical jet luminosity. 
When $\tilde{L} \ll 1$,  the head velocity is non-relativistic, and the jet is collimated by the cocoon pressure. Otherwise the head propagates freely with a relativistic velocity, and the cocoon is unable to collimate the jet.
The right panel of Fig.~\ref{fig: density_and_Ltilde}  depicts $\tilde{L}$ for $L_\jet=10^{50}~\erg/\s$. 
The white region represents the location of the transition from uncollimated to collimated regimes. 
The polar cavities (the red regions in the right panel of Fig.~\ref{fig: density_and_Ltilde}) define regions where the jet is free to propagate conically due to the low ambient density and pressure, whereas in blue regions, only a slowly expanding bow shock is possible. One can expect that, in this case, a small bubble inflates in the very vicinity of the central source, and because  the expansion velocity is small in most directions due to a high ambient density, the flow within the bubble is redirected towards the axis where a free escape is possible. In such a way, a  jet is formed along the axial funnel. The flow freely expands within the funnel until the pressure within the flow is balanced by the pressure of the ambient medium. The shape of the jet and the flow parameters are then found as follows.

The parameters of the outflow are determined by the pressure of the confining medium, $p_\mathrm{ext}(R,z)$ \citep{komissarov_magnetic_2009, lyubarsky_asymptotic_2009}.   The boundary of the jet is determined by the pressure balance condition
\begin{equation}
   \frac{B'^2}{8 \pi} = p_\mathrm{ext}(R,z) \ ,
\end{equation}
where $B'$ is the magnetic field in the comoving frame.
Taking into account that $B_{\phi}/B_\varrho=R/R_\Light$ and that $B'_\varrho=B_\varrho$, $B'_{\phi}=B_{\phi}/\gamma$, this relation is reduced to
\begin{equation}
\label{eq: equilibrium}
    \left(\frac{RB_\varrho}{R_\Light\gamma}\right)^2=8\pi p_\mathrm{ext}(R,z) \ .
\end{equation}
Plugging Eqs.~(\ref{eq: Bp}) and~(\ref{eq: luminosity}) into Eq.~(\ref{eq: equilibrium}) we get the condition
\begin{equation}
\label{eq: equilibrium2}
    \frac{L_\jet}{2 \pi c R^2 \gamma^2} = p_\mathrm{ext}(R,z) \ . 
\end{equation}
At the condition
\begin{equation}
\label{eq: causality}
    \frac{R}{z} \ll \frac{1}{\gamma} \ ,
\end{equation}
the magnetic structure locally relaxes to transverse equilibrium such that $B'_{\phi}\sim B'_\varrho$ at each distance from the origin. On account of Eqs.~(\ref{eq: Bp}) and (\ref{eq: Bphi}) this implies
\begin{equation}
\label{eq: gamma_eq}
    \gamma= \gamma_1 \equiv \frac{R}{R_\Light}.
\end{equation}
Then the boundary condition Eq.~(\ref{eq: equilibrium2}) is reduced to an equation for the shape of the jet:
\begin{equation}\label{jetshape}
    \frac{L_\jet R_\Light^2}{2 \pi R^4 c} = p_\mathrm{ext}(R,z) \ .
\end{equation}
Plugging typical parameters for the jet and assuming a light cylinder radius of few Schwarzschild radii we find
\begin{equation}
\label{eq: pressure}
    L_{51} R^2_{L,6} = 1.6 \times 10^{31} R_5^4~  p_{\mathrm{ext}} (R,z) \ , 
\end{equation}
where $L_{51} = 10^{51}~\erg /\s $, $R_{L,6} = 10^{6}~\cm $, and $R_{5} = 10^{5}~\cm $.

At the condition opposite to Eq.~(\ref{eq: causality}), the azimuthal magnetic field dominates even in the comoving frame, and the Lorentz factor of the flow is determined by the curvature of the magnetic surface:
\begin{equation}
    \gamma=\gamma_2 \equiv \left(-\frac R3\frac{\de ^2R}{\de z^2}\right)^{-1/2} \ .
\end{equation}
In this case the flow is accelerated slower than in the equilibrium case.
Specifically, for a power law shape of the jet, 
\begin{equation}
\label{eq: power_law_shape}
    \frac R{R_\Light}=C\left(\frac{z}{R_\Light}\right)^k \ ,
\end{equation}
one gets
\begin{equation}
\label{eq: gamma_noneq}
    \gamma_2 = \frac{ \sqrt{3}}{C\sqrt{k(1-k)}}\left(\frac{z}{R_\Light}\right)^{1-k} = \frac{ \sqrt{3}}{\sqrt{k(1-k)}}\left(\frac {z}{R}\right) \ .
\end{equation}
Equations (\ref{eq: gamma_eq}) and (\ref{eq: gamma_noneq}) describe two asymptotic limits. In the intermediate zone, one can just take a smaller value of $\gamma$ from those provided by these two formulas.  

%%%%%%%%%%%%%%%%%%%%%%%%%%%%%%%%%%%%%%%%%%%%

\section{Results} 
\label{sec: results}

Figure~\ref{fig: density_and_Ltilde} shows a 2D slice of the original simulation data  on the $x$-$z$ plane at the time $t =1~\s$. 
The left panel depicts the density structure of the central region superposed onto the velocity field lines which expand roughly outwards radially from the center. 
The density is maximal along the equator as a disk has formed and a strong wind emerges from it. 
At the same time due to the conservation of the angular momentum, the density drops dramatically in the polar regions creating two cavities where the $z$-component of the velocity field is inverted resulting in polar inflows. 
The polar inflows for this ($t =1~\s$) snapshot terminates at roughly $z \simeq 5000~\km$, above which the velocity field is directed outward. 
The low ejecta density (and ejecta pressure) of the polar cavities suggests a favorable environment to harbor a conically shaped jet.  

We analyze the 2D $(R,z)$ slices of the 3D simulation box, assuming for simplicity cylindrical symmetry. As the left panel of Fig.~\ref{fig: density_and_Ltilde} illustrates, this assumption is reasonable. 
Within the 2D slices we merged the different resolution layers of the simulation into a single one to maintain as much details as possible at every scale. 
The $z$ and $x$ axes divide the 2D slice of the simulation in four quadrants.
When we impose the axial-symmetry and equatorial symmetry of the system (the northern and southern hemispheres are qualitatively the same),
in order to reduce the noise of the collimation contours, we perform a simple average our physical quantities among the four quadrants.

The outer boundary of the jet is determined by the pressure balance Eq.~(\ref{eq: equilibrium2}). 
At the condition~(\ref{eq: causality}), the shape of the jet could be directly found from Eq.~(\ref{eq: pressure}).
The outflow is produced by the Blandford-Znajek mechanism \citep{blandford_electromagnetic_1977}, according to which the angular velocity of the magnetosphere,  $\Omega$, is one half of the angular velocity of the horizon, 
\begin{equation}
    \Omega_\mathrm{H}=\frac{c}{R_\mathrm{S}}\left(\frac{a}{1+\sqrt{1-a^2}}\right) \ ,
    \end{equation}
where $R_\mathrm{S}$ is the Schwarzschild radius of the black hole, and $a$ is its rotational parameter.
 The corresponding light cylinder radius of the resulting $2.5~\mathrm{M}_\odot$  black hole with  $a \approx 0.65$ equals $R_\Light =c/\Omega= 2 c /\Omega_\mathrm{H} =  40~\km$.
With this $R_\Light$ and the luminosity range that we consider 
$\gamma$, determined by Eq.~(\ref{eq: gamma_eq}), is smaller than the value obtained by Eq.~(\ref{eq: gamma_noneq}). 

\begin{figure}
    \centering
    \includegraphics[scale=0.3]{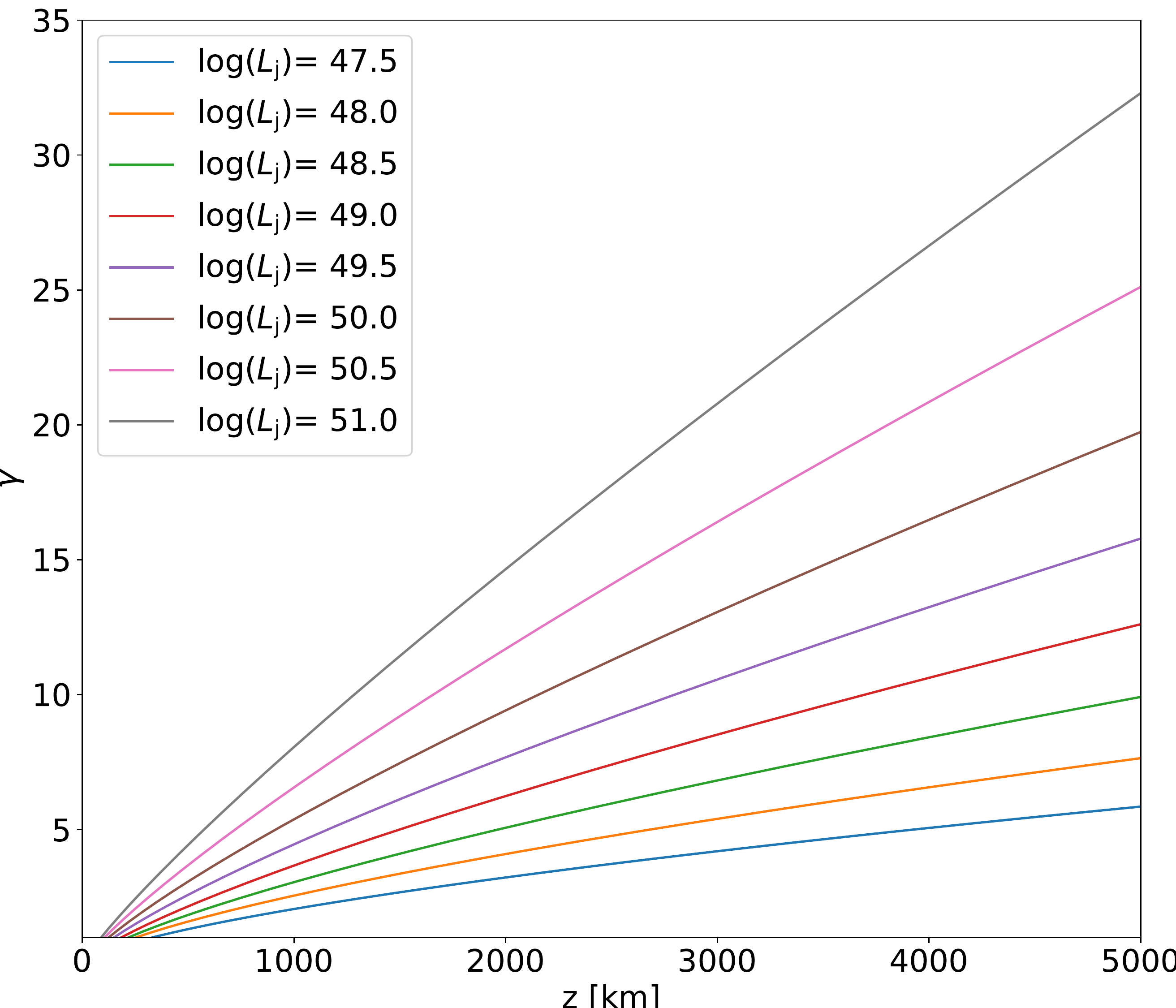}
    \caption{The Lorentz factor as a function of the height, $z$, for different  jet luminosities. The jet luminosity values are the same as  shown in Fig.~\ref{fig: pressure_and_slope}. }
    \label{fig: gamma_factor}
\end{figure}

We first estimate the collimation contour using Eq.~(\ref{eq: pressure}), assuming $\gamma = \gamma_1$ from Eq.~(\ref{eq: gamma_eq}).
Then we calculate the fit for the contour assuming Eq.~(\ref{eq: power_law_shape}) as the shape of the jet. 
At this point we have both the pre-factor $C$ and the slope $k$ and we determine $\gamma = \gamma_2$ from Eq.~(\ref{eq: gamma_noneq}).
Along this fit we calculate the radial position $R_\crit$ where $\gamma_2 = \gamma_1$ and we use the resulting $\gamma$ up to this radius.  We extrapolate the results from $R<R_\crit$ to  $R>R_\crit$ values, but this extrapolation should serve only as a rough indication to the real result in this regime. 
Figure~\ref{fig: gamma_factor} shows how the flow accelerates with height for different values of the luminosity. 

Figure~\ref{fig: pressure_and_slope} depicts the equilibrium shape of the jet for different values of the jet luminosity.
Naturally, the jet opening region widens when the luminosity increases.
Also shown in the background (left panel) the ambient pressure of the ejecta, which is maximal at the equator and minimal along the polar cavities. 
The corresponding fits to the collimation contours are shown on the right panel.
For each fitting curve we measured the local slope and we calculate the corresponding angle of the tangent, which translates to the opening angle of the jet. The color scale of the fitting curves corresponds to the local value inferred for the opening angle.

As the luminosity increases the opening angle corresponding to the collimation contour increases monotonically from $1.7^\circ$ (for $L_\jet \simeq 10^{47}~\erg/\s$) up to $12^\circ$ (for $L_\jet \simeq 10^{51}~\erg/\s $) at $z_\s= 5000~\km$ . 
These data are summarized in Table~\ref{table} where we report the luminosity, the corresponding opening angle and the Lorentz factor $\gamma_{\s}$ at $z_\s = 5000~\km$. 
We also provide the isotropic equivalent luminosity, $L_{\iso,\jet,\s}$, at this stage.
It is important to note that $\gamma_{\s}$ is not the final Lorentz factor as the outflow will accelerate and  expand sideways later. 
The value of $\gamma_{\s}$ also adds a correction factor of the order of $\theta_{\gamma_\s} \sim 1/\gamma_{\s}$ to the final opening angle of the jet. 
Also shown is $\hat L_{\iso,\jet} \equiv 4 L_\jet / (\theta_\jet+\theta_{\gamma_\s})^2 $, the effective isotropic equivalent luminosity corresponding to $\theta_\jet+  1/\gamma_{\s}$.

Using the collimation contour and varying the luminosity of the jet $L_\jet$, we are able to determine the relation between the luminosity and a desired opening angle $\theta_\jet$ of the jet.
The corresponding energy $E_{\jet,\mathrm{coll}}$ to the selected luminosity is simply obtained multiplying $L_\jet$ by the jet engine working time $t_\engine$ (which is of the order of seconds or less for sGRBs).
This is also crucial to determine whether $E_{\jet,\mathrm{coll}}$ is bigger or smaller than the energy $E_{\jet,\esc}$ required by a jet to break out from the ejecta.

\begin{table}
\vskip.5cm
\begin{center}
\begin{tabular}{|c|c|c|c|c|c|}
\hline 
$\log(L_\jet)$ & $\theta_\jet$ & $\log(L_{\iso,\jet,\s})^{~ \dagger}$ & $\gamma_\s^{~ \dagger} $ & $\theta_\jet + \theta_{\gamma_\s}$ & $\log(\hat L_{\iso,\jet})$ \\
$[\erg/\s]$ &  $[\deg]$ & $[\erg/\s]$ & $ $ &  $[\deg]$ & $[\erg/\s]$\\
\hline 
\hline
47.5 & 1.7 & 51.1 & 5 & 11.5 & 49.5 \\
\hline
48.0 & 2.4 & 51.4 & 7 & 9.9 & 50.1 \\
\hline
48.5 & 3.3 & 51.6 & 9 & 9.1 & 50.7 \\
\hline
49.0 & 4.4 & 51.8 & 12 & 9.0 & 51.2 \\
\hline
49.5 & 5.7 & 52.1 & 15 & 9.3 & 51.7 \\
\hline
50.0 & 7.3 & 52.4 & 19 & 10.2 & 52.1 \\
\hline
50.5 & 9.5 & 52.7 & 25 & 11.8 & 52.5 \\
\hline
51.0 & 12.6 & 52.9 & 32 & 14.3 & 52.8 \\
\hline
\end{tabular} 
\end{center}
\caption{
Summary of the parameters calculated for the collimation contours for different luminosities (Fig.~\ref{fig: pressure_and_slope}). The angle $\theta_\jet$, the corresponding isotropic equivalent luminosity $L_{\iso,\jet,\s} = 4 L_\jet / \theta_\jet^2 $ and the Lorentz factor $\gamma_{\s}$ are the parameters at the end of the calculations, that is at $z_\s=5000~\km$. 
The  Lorentz factor continues to increase after the jet emerges from the ejecta {and the thermal energy of the jet may cause it to expand as well. } In the second-last column we added a correction factor {for the expansion}, $\theta_{\gamma_\s} \sim 1/\gamma_\s$, to the angle. 
In the last column we show the isotropic equivalent luminosity $\hat L_{\iso,\jet} \equiv 4 L_\jet / (\theta_\jet+\theta_{\gamma_\s})^2 $ which accounts for this corrected angle.
\newline
\hspace*{2.5mm} $\dagger$ \footnotesize{These are not the final Lorentz factor and the final isotropic equivalent luminosity as the jet will accelerate and expand after it breaks out from the ejecta. This is particularly important for the low $\gamma_\s$ cases. } }
\label{table}
\end{table}

\begin{figure*}
    \centering
    \includegraphics[scale=0.43]{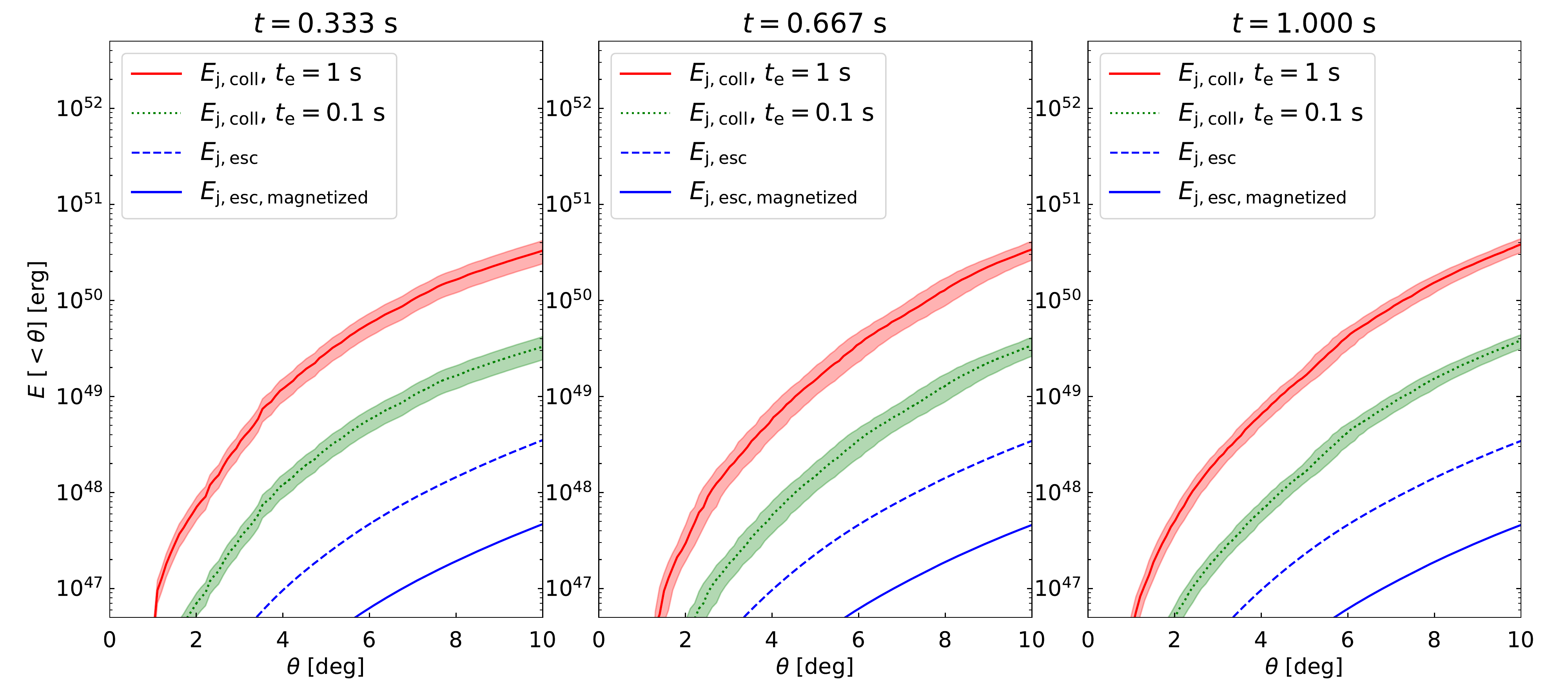}
    \caption{Escape energy, $E_{\jet,\esc}$, (blue lines) and collimation energy, $E_{\jet,\mathrm{coll}}$,  (red and green lines) versus $\theta$ for three different snapshots at: $t=t_1=0.33~\s$ (left), $t =2 t_1 = 0.667~\s$ (center), and $t = 3 t_1 = 1~\s$ (right). 
    The red (and green lines) are drawn varying the luminosity of the collimation contour and calculating, from the contour fit, the corresponding opening angle $\theta$ measured at $z_s=1670~\km$ (left), $z_s=3300~\km$ (centre), and $z_s=5000~\km$ (right), which approximately represent the maximal heights of the low-density, low-pressure cavity at each snapshot, respectively. 
    The jet energy is calculated from the luminosity assuming a jet engine working time of $1~\s$ (red solid line) or $0.1~\s$ (red dotted line). 
    The red-shaded region represents the uncertainty on the estimate taking different values of the height at which the slope is measured. 
    The blue lines indicate the energy required for a magnetic (solid) and hydrodynamic (dashed) jets to overrun the expanding ejecta and escape  the external vacuum as expressed by Eq.~(\ref{eq: escape}). 
    Collimated jets that break out should be on red (green) lines and above the blue lines. }
    \label{fig: energy_within}
\end{figure*}

Figure~\ref{fig: energy_within} describes  $E_{\jet,\mathrm{coll}}$ and $E_{\jet,\esc}$ versus the azimuthal angle $\theta_\jet$ for different snapshot at: $t=0.333~\s$, $t=0.667~\s$ and $t=1~\s$. 
The last one corresponds to the snapshot used in the previous analysis. 
For the collimation energy we assumed two different values for the jet engine working time: $t_\engine=1~\s$ and $t_\engine = 0.1~\s$. 

The uncertainty in the collimation energy at a given angle  is calculated varying  the height from 3/4 to 5/4 of the chosen reference height at that given time. 
The required escape energy is calculated for hydrodynamic and magnetic jets. 

The plots have a simple interpretation: collimated jets that manage to break out from the ejecta have to be on the  $E_{\jet,\mathrm{coll}}$ line and above $E_{\jet,\esc}$. 
As the former is  above the latter we expect that in the relevant energy range of $10^{47}$--$10^{50}$ erg jets will break out and will be collimated by the ejecta. 

\begin{figure}
    \centering
    \includegraphics[scale=0.30]{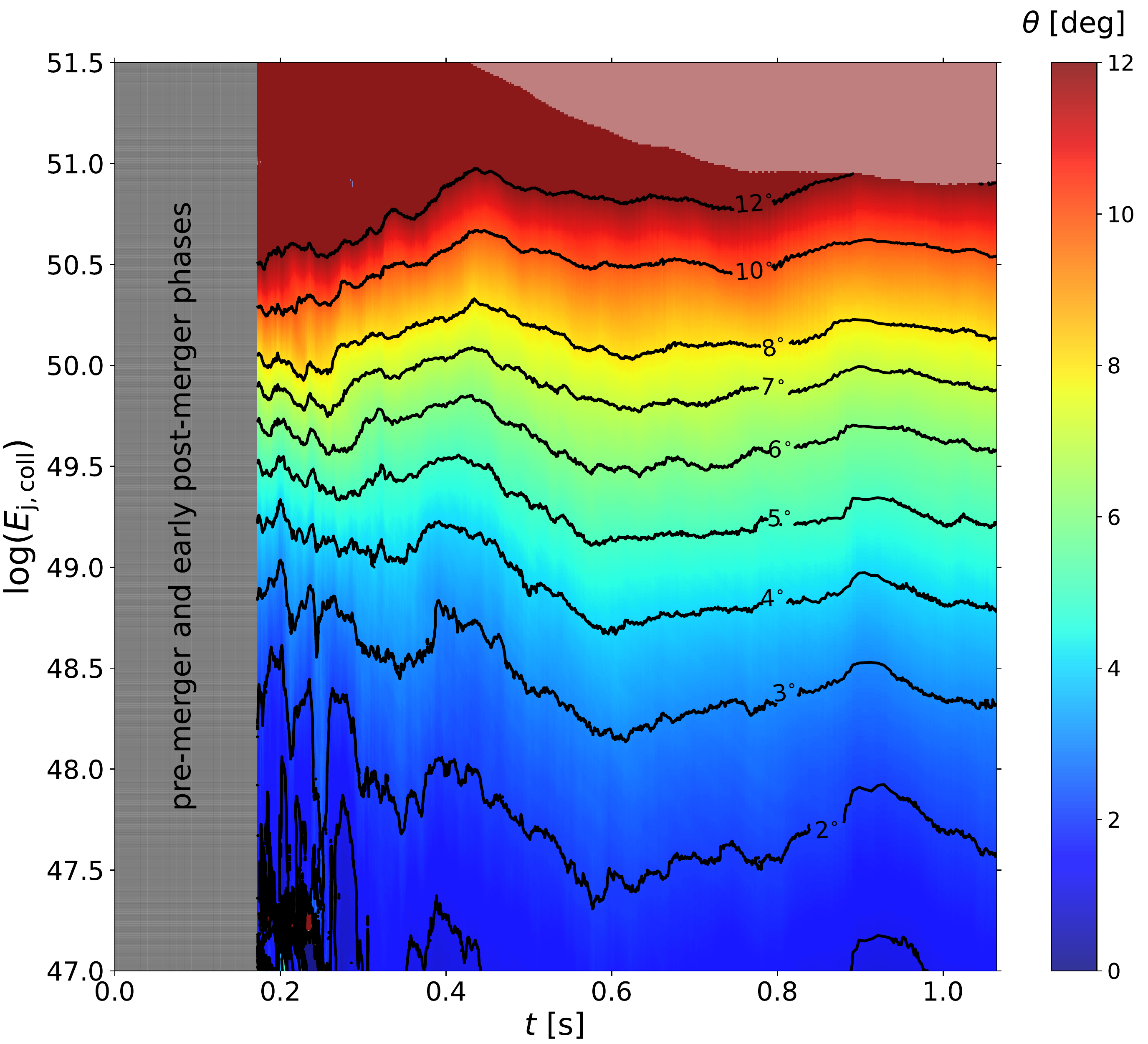}
    \caption{Evolution of the angle $\theta$ as a function of the time $t$ and the collimation energy $E_\mathrm{j, coll}$ (in units of $\erg$) for $t_\engine=1~\s$.
    The gray area at the left of the plot represents the pre-merger and early post-merger phase of the system, where the calculation of the collimation contour is still very noisy.
    The fainter area on the top of the map represents the region where the contour is no more calculated assuming $\gamma = R/R_\Light$ but using the curvature of the jet magnetic surface. 
    The black curves represent the contour lines for fixed values of $\theta$, from $1^\circ$ to $10^\circ$. 
    The color scale on the right indicates the value of $\theta$ in degrees. }
    \label{fig: energy_vs_angle_vs_time}
\end{figure}

The similarity between the three panels of Fig.~\ref{fig: energy_within}, and in particular,  between the two later snapshots suggests that further evolution of the ejecta would yield comparable results. 
This point is explored further in
Fig.~\ref{fig: energy_vs_angle_vs_time} where the collimation angles are shown over time and as a function of the collimation energy. 
We tracked the evolution from $t=0.2\,\s$ to $t=1.06\,\s$, roughly 1 second after the merger. 
The early pre-merger and early post-coalescing phase are marked with a grey area, indicating that the calculation of the equilibrium contours in this phase is meaningless (pre-merging) or too noisy (early post-merger).
The black curves in the plot represent the contour lines for fixed values of $\theta$, from $1^\circ$ to $12^\circ$. 
After an initial rise we see that around $t\simeq 0.6\,\s$ the curves begins to stabilize around constant values.
Close to the end of the simulation an energy of $ E_{\jet,\mathrm{coll}} \sim 10^{49}~\erg $ roughly corresponds to a collimation angle of $ \theta \simeq 4.5^\circ $. 

\begin{table}
\vskip.5cm
\begin{center}
\begin{tabular}{|c|c|c|c|c|}
\hline 
\backslashbox[28mm]{ $t_\mathrm{e}$}{$t_\mathrm{d}$} & 0.01 ~s& 0.1~s & 0.5~s & 1~s \\ 
\hline 
0.1~s & $\sim 2$ & 3 & 27 & 100 \\ 
\hline 
0.5~s & $\sim 2$ & $\sim 2$ & 3 & 6 \\ 
\hline 
1~s & $\sim 2$ & $\sim 2$ & 2.25 & 3 \\ 
\hline 
\end{tabular} 
\end{center}
\caption{Values of the pre-factor $[(t_\mathrm{d}/t_\engine)^2+2]$ in Eq.~(\ref{eq: escape}) as a function of the delay time $t_\mathrm{d}$ between the merger time and the jet launch and the engine working time $t_\engine$ of the launched jet.}
\label{table: pre-factor}
\end{table}

In the presented calculation we considered a negligible delay between the merger and the jet launch ($t_\mathrm{d} = 0$). 
Table~\ref{table: pre-factor} summarizes how the factor $[(t_\mathrm{d}/t_\mathrm{e})^2+2]$ varies as a function of the engine working time $t_\engine$ and the delay time $t_\mathrm{d}$. 
If $t_\mathrm{d} > t_\engine$, for short  engine working times $t_\engine$, the pre-factor may increase dramatically. 
This may result in a choked jet that is unable to break out from the expanding ejecta.
This effect is negligible for $t_\mathrm{d} < t_\engine$ .

%%%%%%%%%%%%%%%%%%%%%%%%%%%%%%%%%%%%%%%%%%%%%%%%%%

\section{ Implications to GW170817 and other sGRBs} \label{sec:implications}

We begin by considering the implications of these results to the observations of GW170817 and its EM counterparts. 
As mentioned earlier the multi-messenger observations of GW170817 allowed to put constraints on the general properties of the jet harbored in this BNS merger.
The afterglow light curve in combination with the superluminal motion of the jet core \citep[e.g.][]{mooley_superluminal_2018} finds a powerful ($E_{\jet} = 10^{49-50}~\erg$)  narrow cone,  ($\theta < 5^\circ$) jet. 
This value is much higher than the required escape energy  which is around $\sim 10^{47}$--$10^{48}$~ergs.
With this energy the $E_{\jet,\mathrm{coll}}$ (see Fig.~\ref{fig: energy_within} and Table~\ref{table}) corresponds to a collimation angle of $\simeq 4^\circ$--$7^\circ$ which is consistent with observations.   
 
The results are also encouraging considering other sGRBs whose $E_{\iso,\jet}$ is typically significantly weaker than the one implied for the jet in GW170817. For example, \citet{guetta_luminosity_2005} and \citep{wanderman_rate_2015} find that typical isotropic equivalent luminosity \footnote{ \citet{guetta_luminosity_2005} and \citep{wanderman_rate_2015} find that the break in the light curve from a shallow to a steep decline is at $L_* \sim 10^{52}~ \erg/\s$.  The average and median luminosities are approximately two orders of magnitude lower. }
of sGRBs is around $10^{50}~\erg/\s$, almost two orders below those at the core of the jet in GW170817. 
Examination of Fig.~\ref{fig: energy_within} and Table~ \ref{table} shows that even much weaker jets could escape and the overall estimated isotropic equivalent luminosity $ \hat{L}_{\iso,\jet} $ is within the range of observed values.

%%%%%%%%%%%%%%%%%%%%%%%%%%%%%%%%%%%%%%%%%%%%%%%%%%

\section{Conclusions} \label{sec:conclusions}

We analyzed the data for the merger remnant of a recent numerical general relativistic MHD simulation of a binary neutron star merger with masses of $1.2 ~\mathrm{M}_\odot $ and $1.5~ \mathrm{M}_\odot$ \citep{kiuchi_self-consistent_2022}. 
Running up to $1~\s$ after the merger this numerical simulation is the longest carried out so far and with a realistic equation of state, MHD, and neutrino physics it is able to yield a good estimate of the resulting mass ejection. 
In our analysis we ask the question which kind of jets can break out from the ejecta and be collimated to a narrow angle at the same time. 
Overall, our analysis suggests that the post-BNS merger ejecta anisotropic structure creates a favorable environment to  successfully collimate  jets in the energy range seen in sGRBs. 
This arises because of the almost empty cavity (see Fig.~\ref{fig: density_and_Ltilde}) that forms, due to the conservation of the angular momentum, around the rotation axis. 
It is important to note that the same jets cannot escape in any other direction within the ejecta. 
This fact emphasizes the importance of the high precision full numerical simulations. 
Jets that broke out through the cavity would have been choked  if the surrounding ejecta had been approximated as spherically symmetric.  
Another important factor is that the engine operation time cannot be too short as compared to the time delay between the onset of the jet and the merger (see Table~\ref{table: pre-factor} and \cite{gottlieb_propagation_2022}.

Previous works showed that jets with a  long engine time ($t_\mathrm{e} \sim 1~\s$) and similar luminosities to this work ($L_{\jet} \sim 10^{50}~\erg/\s$) can pierce merger ejecta in both idealised (isotropically distributed) simulations \citep{lazzati_off-axis_2017} and more complex setups originated from GRMHD simulations  \citep{lazzati_two_2021,pavan_short_2021}, both based on the setups of \citet{ciolfi_first_2019}).
However, in those works the jet was injected at  a few hundred $\km$ from the compact source (roughly at 1/3 of the ejecta maximum radius in their setup), which circumvents the problem to carve through the densest near-isotropic ejecta at the center
and thus making a direct comparison with our prediction harder.

Our estimates of the luminosity (or energy) of jets  that can be both collimated and break out from the ejecta are compatible with observations of GW170817 and its EM counterparts. 
We find that a jet with energy of $\sim 10^{49-50}\,\erg$, will be collimated, by the outflow found in the numerical simulation to $\sim 4^\circ-7^\circ$ just like the observations suggest. 
We also find that the range of isotropic equivalent luminosities observed in sGRBs is compatible with the range that we find for jets that can be collimated and break out from the ejecta. 

%%%%%%%%%%%%%%%%%%%%%%%%%%%%%%%%%%%%%%%%%%%%%%%%%%

\section{Acknowledgements}
We thank our anonymous referee for a constructive and insightful report. This work was supported by the ERC grants TReX and Multijets
(TP, YL and MP) and by Grant-in-Aid for Scientific Research (Grant No.~JP20H00158) of Japanese MEXT/JSPS (KK and MS). 
This work used computational resources of the supercomputer Fugaku provided by RIKEN through the HPCI System Research Project (Project ID: hp220174). 
The simulation was also performed on Sakura, Cobra, and Raven clusters at the Max Planck Computing and Data Facility and on the Cray XC50 at CfCA of the National Astronomical Observatory of Japan.

\bibliography{file}{}
\bibliographystyle{aasjournal}

\end{document}